\documentclass[11pt,a4paper]{article}
\pdfoutput=1
\usepackage{jheppub}

\usepackage{color,epsfig} 
\usepackage{ifpdf}
\usepackage{amsmath} 
\usepackage{amssymb} 
\usepackage{bm}

\usepackage[english]{babel}
\usepackage{slashed}
\usepackage{multirow}

\usepackage{subfigure}

\title{Cornering Scalar Leptoquarks at LHC}

\author[a]{Ilja Dor\v sner,} 
\author[b,c]{Svjetlana Fajfer} 
\author[b]{and Admir Greljo} 

\affiliation[a]{University of Split, Faculty of Electrical Engineering, Mechanical Engineering and Naval Architecture in Split (FESB), R.\ Bo\v skovi\' ca 32, 21000 Split, Croatia}
\affiliation[b]{J.\ Stefan Institute, Jamova 39, P.\ O.\ Box 3000, 1001
  Ljubljana, Slovenia}
\affiliation[c]{Department of Physics,
  University of Ljubljana, Jadranska 19, 1000 Ljubljana, Slovenia}

\emailAdd{dorsner@fesb.hr}
\emailAdd{svjetlana.fajfer@ijs.si} 
\emailAdd{admir.greljo@ijs.si} 

\abstract{We study implications of large lepton-quark-leptoquark couplings for direct leptoquark searches at Large Hadron Collider. We present all existing flavor constraints on the strength of these couplings assuming that leptoquarks under consideration interact exclusively with charged leptons and quarks of the same generation. We find that these leptoquarks can have sizeable couplings to the Standard Model fermions. This insures a self consistency of our study. We discuss the leptoquark production mechanisms at LHC and demonstrate the importance of inclusion of a $t$-channel pair production and, in particular, a single leptoquark production through a recast of an existing CMS search at LHC for the second generation leptoquark. Our recast yields the best direct limit on Yukawa coupling of the second generation leptoquark that couples to a muon and a strange quark to date.}

\arxivnumber{1406.4831}

\begin{document}

\maketitle

\section{Introduction}
\label{sec:intro}

A large number of well-motivated models that go beyond the Standard Model (SM) of elementary particle physics predicts existence of leptoquarks~\cite{Pati:1974yy,Georgi:1974sy}. These hypothetical particles make leptons couple directly to quarks and vice versa. The leptoquark discovery would thus signal the matter field unification. This, on the other hand, would nicely dovetail with the observed unification of weak and electromagnetic interactions.

The leptoquark properties have been extensively studied in literature. It is commonly accepted that they come in ten different multiplets~\cite{Buchmuller:1986zs} under the SM gauge group of $SU(3) \times SU(2) \times U(1)$. One half of these multiplets is of scalar nature and the other half is of vector nature under the Lorentz transformations. In this note we investigate only scalar leptoquarks. 

There are leptoquarks that can destabilize the matter if one allows for presence of all $SU(3) \times SU(2) \times U(1)$ invariant tree level contractions between these fields and the SM fermions. The current experimental limits on proton decay put severe constraints on the strength of these contractions and/or the masses of associated leptoquarks. We accordingly focus in our study on those leptoquark multiplets that do not contribute to nucleon decay at tree level.

All existing accelerator searches for light leptoquarks are assumption driven. These assumptions can and should be scrutinized for potential flaws. Here we focus on one particular assumption we find troublesome. Namely, it is commonly assumed that the pair production of leptoquarks is purely QCD driven. Note, however, that the sizeable Yukawa couplings of the leptoquarks with the SM fermions could influence pair production as we demonstrate later on. This regime would also make single leptoquark production very relevant at hadron colliders~\cite{Hewett:1987yg,Eboli:1987vb,DeMontigny:1989yd,Ohnemus:1994xf,Eboli:1999ye,Belyaev:2005ew}. Again, existing experimental studies do not address this part of parameter space.

We study implications of large Yukawa couplings for a pair production and a single production of leptoquarks at Large Hadron Collider (LHC). We accomplish this through a recast of an existing CMS search for the second generation leptoquark. To be self consistent we show that the current flavor constraints do not exclude the parameter space we are interested in. Interestingly enough, our recast shows that the current LHC data place stringent limits on the mass of scalar leptoquark that has large couplings to the SM fermions. We find these limits to be more relevant than the corresponding limits one infers from the flavor physics measurements.

Our work is organised as follows. In section~\ref{sec:framework} we present two leptoquark multiplets we study and discuss their couplings to the SM fermions. Leptoquark production mechanisms at LHC are discussed in section~\ref{sec:production}. We provide relevant flavor physics constraints on the scalar leptoquark Yukawa couplings in section~\ref{sec:flavor}. The recast of the search for the second generation leptoquark is given in section~\ref{sec:recast}. We finally conclude in section~\ref{sec:conclusions}.

\section{Framework}
\label{sec:framework}

The scalar leptoquark multiplets one usually finds listed in literature~\cite{Buchmuller:1986zs} comprise $(\overline{\mathbf{3}},\mathbf{1},1/3)$, $(\overline{\mathbf{3}},\mathbf{1},4/3)$, $(\overline{\mathbf{3}},\mathbf{3},1/3)$, $(\mathbf{3},\mathbf{2},7/6)$ and $(\mathbf{3},\mathbf{2},1/6)$, where we opt to denote leptoquarks via their transformation properties under the SM gauge group of $SU(3) \times SU(2) \times U(1)$. Our normalization is such that $Q=I_3+Y$, where $Q$ is the electric charge, $I_3$ stands for appropriate eigenvalue of the diagonal generator of $SU(2)$, and $Y$ represents the $U(1)$ (hyper)charge. This classification implicitly assumes Majorana nature of neutrinos and/or kinematical inaccessibility of right-handed neutrinos. If one departs from these assumptions there is one more multiplet --- $(\overline{\mathbf{3}},\mathbf{1},-2/3)$ --- that should be added to the scalar leptoquark list~\cite{Dorsner:2012nq}.

The list of leptoquarks that could be potentially light is somewhat smaller in view of the current data on matter stability. Namely, there are only two scalar leptoquark multiplets --- $(\mathbf{3},\mathbf{2},7/6)$ and $(\mathbf{3},\mathbf{2},1/6)$ --- that are not dangerous for proton decay at tree level. These two multiplets can thus have sizeable Yukawa couplings to matter and be light enough to be accessible in accelerator searches. This is not to say that other scalar leptoquarks could not be light and have non negligible couplings. It is just that such scenarios require additional assumptions and/or additional supporting structures in order to be viable.

We, for definiteness, consider the SM that is extended with a single scalar leptoquark (LQ) representation. This LQ, for aforementioned reasons, we take to be either $\tilde{R}_2\equiv(\mathbf{3},\mathbf{2},1/6)$ or $R_2\equiv(\mathbf{3},\mathbf{2},7/6)$. (Here, we also use notation for the LQ states that was introduced in ref.~\cite{Buchmuller:1986zs}.) In fact, to drive our point, we will mainly refer to a $Q=2/3$ component in $\tilde{R}_2$ and a $Q=5/3$ component in $R_2$ in this study. (Note that there is an LQ component in $R_2$ with $Q=2/3$. Our analysis of the $\tilde{R}_2$ component with the same electric charge will be applicable to both of these leptoquarks when we discuss accelerator signatures.)

We now briefly summarize the LQ couplings of $\tilde{R}_2$ and $R_2$ to the SM fermions. A particular ansatz we introduce helps us to simplify our discussion and to perform self consistent recast of accelerator signatures we need to drive our point.

\subsection{The \texorpdfstring{$(\mathbf{3},\mathbf{2},1/6)$}{(3,2,1/6)} case}

The only renormalizable term that describes interactions of $\tilde{R}_2$ with matter is given by
\begin{equation}
\mathcal{L}_\mathrm{Y}=-y_{ij}\bar{d}_{R}^{i}\tilde{R}_{2}^{a}\epsilon^{ab}L_{L}^{j,b}+\textrm{h.c.},
\end{equation}
where we explicitly show flavor indices $i,j=1,2,3$, and $SU(2)$ indices $a,b=1,2$. $y_{ij}$ are elements of an arbitrary complex $3 \times 3$ Yukawa coupling matrix. After expanding $SU(2)$ indices, we obtain
\begin{equation}
\label{eq:main_1}
\mathcal{L}_\mathrm{Y}=-y_{ij}\bar{d}_{R}^{i}e_{L}^{j}\tilde{R}_{2}^{2/3}+(y V_\mathrm{PMNS})_{ij}\bar{d}_{R}^{i}\nu_{L}^{j}\tilde{R}_{2}^{-1/3}+\textrm{h.c.},
\end{equation}
where the LQ superscript denotes electric charge of a given $SU(2)$ doublet component, and $V_\mathrm{PMNS}$ represents Pontecorvo-Maki-Nakagawa-Sakata mixing matrix. All fields in eq.~\eqref{eq:main_1} are specified in the mass eigenstate basis.  

We take both components of $\tilde{R}_2$ to be degenerate in mass. (Splitting the mass degeneracy beyond the $W$ boson mass would drastically modify the LQ phenomenology. It would allow for the decays of the heavier of two leptoquarks into a lighter LQ and a $W$ boson. However, such splitting is inconsistent with the electroweak precision measurements as it induces violent corrections to
$T$~parameter~\cite{Davidson:2010uu}.) We furthermore take the following ansatz for Yukawa coupling matrix $y$: $y_{ij}=\delta_{ij}y_i$, $i,j=1,2,3$. $\tilde{R}_2^{2/3}$ thus couples exclusively to a charged lepton and a down-type quark of the same generation. It has, however, non-zero couplings to more than one generation of fermions at a given instant. This represents slight departure from what is usually assumed in the literature. (This is not to say that there exist no studies that investigate this possibility. See, for example, ref.~\cite{Aaron:2011zz}.) When we present our numerical results we take the limit where only one of these couplings is dominant to simplify our discussion. This, on the other hand, brings our study in line with current analyses one finds in literature on a single generation LQ signatures. (We defer discussion of the more general case to future publications.)

Decay width of $\tilde{R}_2^{2/3}$ to a particular decay channel is
\begin{equation}
\Gamma(\tilde{R}_{2}^{2/3}\to d_{i}e_{i}^{+})=\frac{m_\mathrm{LQ}}{16\pi}\left|y_{i}\right|^{2},
\label{eq:width}
\end{equation}
where $m_\mathrm{LQ}$ is the LQ mass. Correspondingly, branching ratios are given by
\begin{equation}
\beta_{i}=\frac{\left|y_{i}\right|^{2}}{\left|y_{1}\right|^{2}+\left|y_{2}\right|^{2}+\left|y_{3}\right|^{2}}, \qquad i=1,2,3.
\end{equation}
We use expressions given in eqs.~\eqref{eq:main_1} and~\eqref{eq:width} in our numerical simulation.

\subsection{The \texorpdfstring{$(\mathbf{3},\mathbf{2},7/6)$}{(3,2,7/6)} case}

Yukawa couplings of $R_2$ to the SM fermions are
\begin{equation}
\label{eq:main_2}
 \mathcal{L}_\mathrm{Y} = z_{ij}\bar{e}_{R}^{i} R_{2}^{a\,*}Q_{L}^{j,a} -y_{ij}\bar{u}_{R}^{i} R_{2}^{a}\epsilon^{ab}L_{L}^{j,b}+\textrm{h.c.},
\end{equation}
where we explicitly show flavor indices $i,j=1,2,3$, and $SU(2)$ indices $a,b=1,2$. $y$ and $z$ in eq.~\eqref{eq:main_2} are {\it a priori} arbitrary complex $3 \times 3$ Yukawa matrices. 
In the mass eigenstate basis we have 
\begin{equation}
  \label{eq:L23}
\mathcal{L}_\mathrm{Y}= z_{ij} \bar{e}^i_R d^j_L R_2^{2/3\,*} +(z V_\mathrm{CKM}^\dagger)_{ij} \bar{e}^i_R 
  u^j_L R_2^{5/3\,*}+  (y
  V_\mathrm{PMNS})_{ij} \bar{u}^i_R 
  \nu^j_L R_2^{2/3}  - y_{ij} \bar{u}^i_R e^j_L R_2^{5/3} + \textrm{h.c.},
\end{equation}
where the LQ superscript denotes electric charge of a given $SU(2)$ doublet component and $V_\mathrm{CKM}$ represents Cabibbo-Kobayashi-Maskawa mixing matrix.  Clearly, both components of $R_2$ have two sets of couplings to the SM fermions. (We take both of these components to be degenerate in mass.) These two sets, on the other hand, are not related through observed mixing matrices. This makes a self consistent analysis of accelerator signatures of $R_2$ rather difficult. In what follows we investigate a particular case of $R_2^{5/3}$ production at LHC when $y_{ij}=\delta_{ij} y_i$ and $z_{ij}=0$, $i,j=1,2,3$. This is more in line with what is usually studied in literature.

\section{Leptoquark production mechanism at LHC}
\label{sec:production}

We want to demonstrate that the leptoquark production at LHC is not driven solely by QCD induced pair production. It can be substantially influenced by the presence of relatively large Yukawa couplings of leptoquarks to the SM fermions. Moreover, if these couplings are taken to be large one also needs to take into consideration a single leptoquark production~\cite{Hewett:1987yg,Eboli:1987vb,DeMontigny:1989yd,Ohnemus:1994xf,Eboli:1999ye,Belyaev:2005ew} and a $t$-channel leptoquark pair production.

\begin{figure}[tbp]
\centering
\includegraphics[scale=0.75]{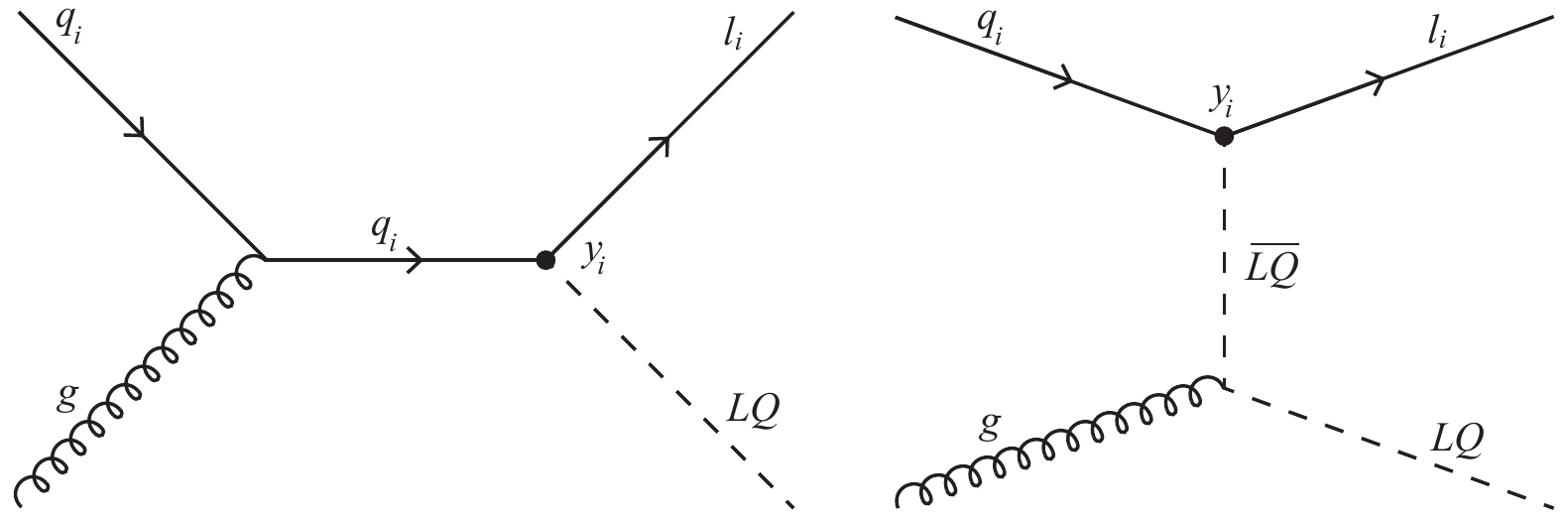}
\caption{\label{fig:-plot-0a} Complete set of the leading-order Feynman diagrams relevant for a single leptoquark production through an $s$-channel (left panel) and a $t$-channel (right panel) at LHC. Here, $y_i$, $i=1,2$, represents Yukawa coupling of a quark $q_i$ ($u$, $d$ and $s$) and a charged lepton $l_i$ ($e$ and $\mu$) with a leptoquark ($LQ$).}
\end{figure}

We first show complete set of Feynman diagrams, at leading-order, that are relevant for a single leptoquark production of $R_2^{5/3}$ and $\tilde{R}_2^{2/3}$ at LHC in figure~\ref{fig:-plot-0a} taking into account our particular ansatz for the couplings of these leptoquarks to the SM fermions. The diagrams shown in figure~\ref{fig:-plot-0a} are an $s$-channel (left panel) and a $t$-channel (right panel). Note that we use generic symbols for all the fields including the leptoquarks in figure~\ref{fig:-plot-0a}. We take into account the composition of a proton and hence refer to $u$ ($d$ and $s$) to account for $R_2^{5/3}$ ($\tilde{R}_2^{2/3}$) production. 

\begin{figure}[tbp]
\centering
\includegraphics[scale=0.75]{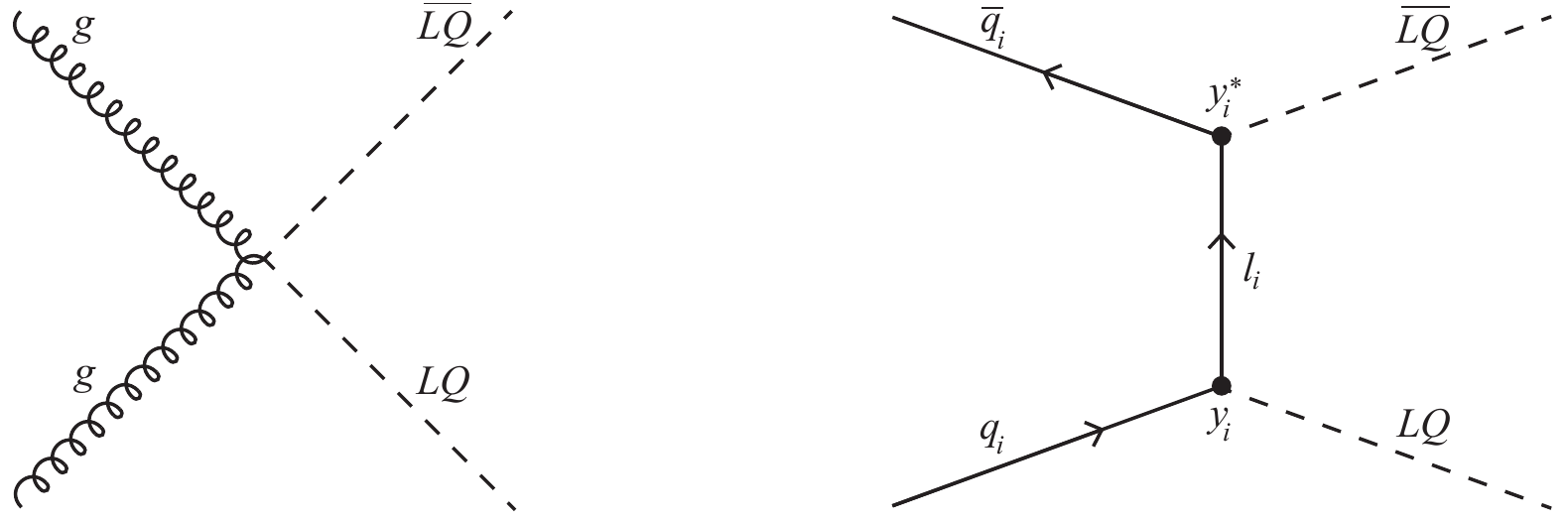}
\caption{\label{fig:-plot-0b} Feynman diagrams relevant for a pair production of leptoquarks at LHC. Representative diagram for a gluon fusion process is shown on the left. The diagram on the right represents a $t$-channel production mechanism. Here, $y_i$, $i=1,2$, represents Yukawa coupling of a quark $q_i$ ($u$, $d$ and $s$) and a charged lepton $l_i$ ($e$ and $\mu$) with a leptoquark ($LQ$).}
\end{figure}

We next show Feynman diagrams that depict the LQ pair production in figure~\ref{fig:-plot-0b}.
The QCD diagrams that contribute to a leptoquark pair production at LHC are numerous. We opt to show a representative diagram for gluon fusion in figure~\ref{fig:-plot-0b} (left panel). There is, on the other hand, only one type of the Yukawa coupling contribution to the leptoquark pair production and it corresponds to a $t$-channel process we show in figure~\ref{fig:-plot-0b} (right panel). The important point to notice is that the amplitude that corresponds to a $t$-channel is proportional to a square of absolute value of the relevant Yukawa coupling. This makes it especially relevant in the limit of large Yukawas. 

One can consider the leptoquark pair production to have three distinct regions. For small Yukawa couplings the total cross section is purely QCD driven. For intermediate Yukawas there exists a region with a negative interference between the QCD diagrams and the $t$-channel diagram, where the total cross section can be decreased by up to 15\% depending on the quark type ($u$, $d$, and $s$), mass of the leptoquark ($m_\mathrm{LQ}$) and strength of the coupling ($y_i$). Finally, there is the region of large Yukawas where the $t$-channel contribution not only dominates the QCD one but significantly enhances the total cross section. We are particularly interested in that region.

In the following, we calculate the leading-order (LO) production cross sections using 
MadGraph~5 (v1.5.11)~\cite{Alwall:2011uj} after we implement 
the LQ models of section~\ref{sec:framework} in FeynRules (v1.6.16)~\cite{Christensen:2008py}. Our calculations are performed for 
$\sqrt{s}=8$\,TeV proton-proton center of mass energy with fixed renormalization 
($\mu_R$) and factorization ($\mu_F$) scales set to $\mu_R=\mu_F=m_\mathrm{LQ}/2$. The cross section 
for the single LQ production takes the following form, 
\begin{equation}
\sigma_{\textrm{single}}(y_i,m_\mathrm{LQ})= a(m_\mathrm{LQ}) |y_i|^2,
\end{equation} 
where the coefficient $a(m_\mathrm{LQ})$ depends on the leptoquark mass but 
not on its coupling to the SM fermions. Therefore, we calculate the cross section for 
$p\,p\to \overline{LQ}\, l_i$ together with $p\,p\to LQ \, \overline{l_i}$ for several $m_\mathrm{LQ}$ choices while setting the coupling $y_i$ to one, i.e, $y_i=1$. We find the functional dependence $a(m_\mathrm{LQ})$ by 
using the appropriate interpolation. Analogously, the cross section 
for the leptoquark pair production is assumed to take the form,
\begin{equation}
\sigma_{\textrm{pair}}(y_i,m_\mathrm{LQ})= a_0(m_\mathrm{LQ})+a_2(m_\mathrm{LQ}) |y_i|^2+a_4 (m_\mathrm{LQ}) |y_i|^4,
\end{equation} 
where the three terms correspond to the QCD pair production, an interference term and
a $t$-channel production, respectively.  In order to obtain the proper functional 
dependence, we calculate the cross section for a given $m_\mathrm{LQ}$ for three values
of the $y_i$ coupling and solve for $a_0(m_\mathrm{LQ})$, $a_2(m_\mathrm{LQ})$ and $a_4(m_\mathrm{LQ})$. We repeat this at several mass
points and use the appropriate interpolation.

\begin{figure}[tbp]
\centering
\includegraphics[scale=0.73]{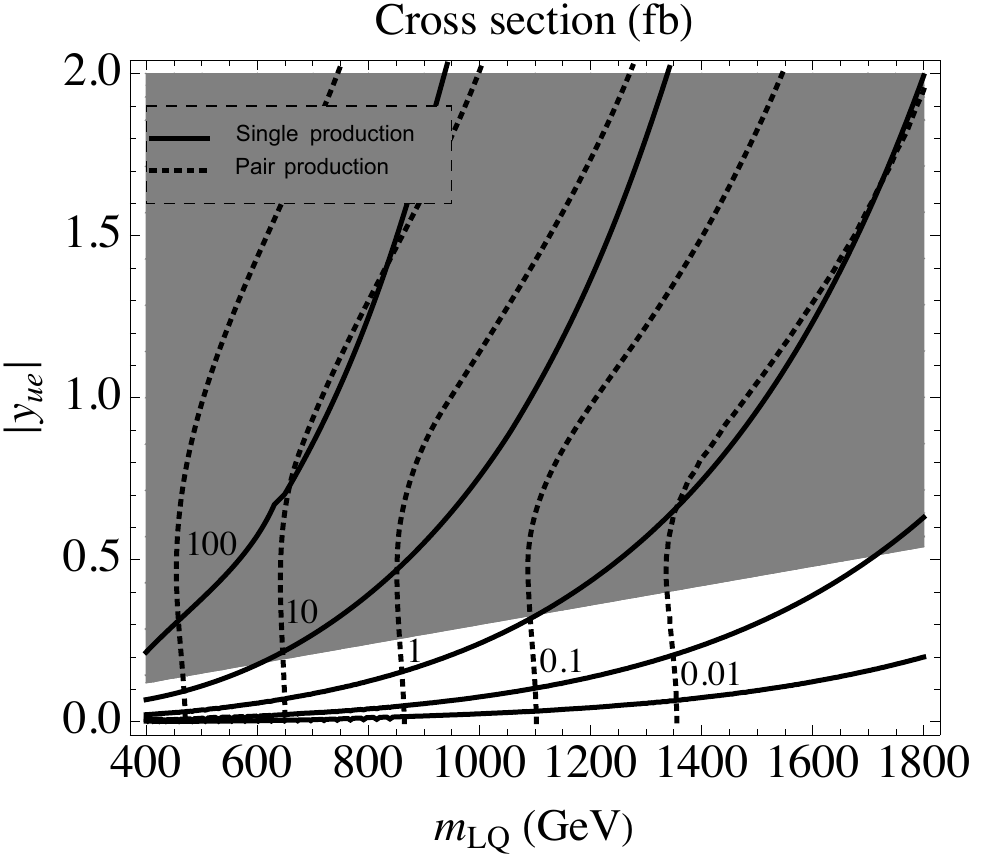}
\includegraphics[scale=0.73]{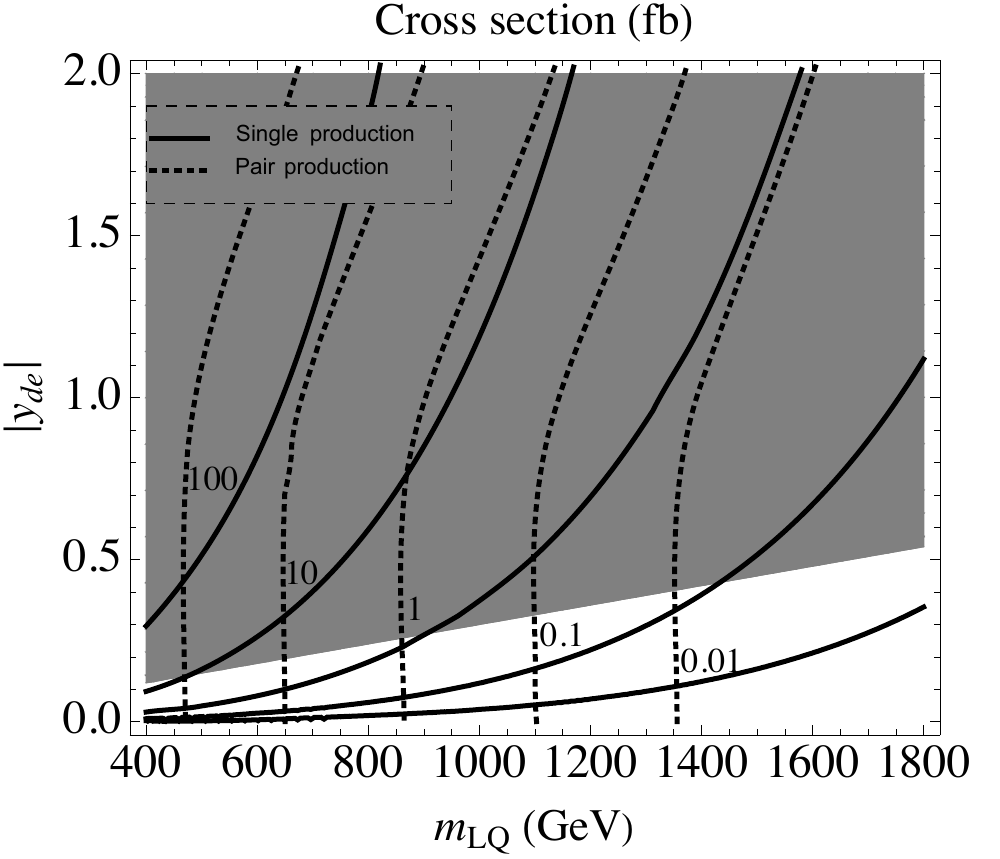}
\includegraphics[scale=0.73]{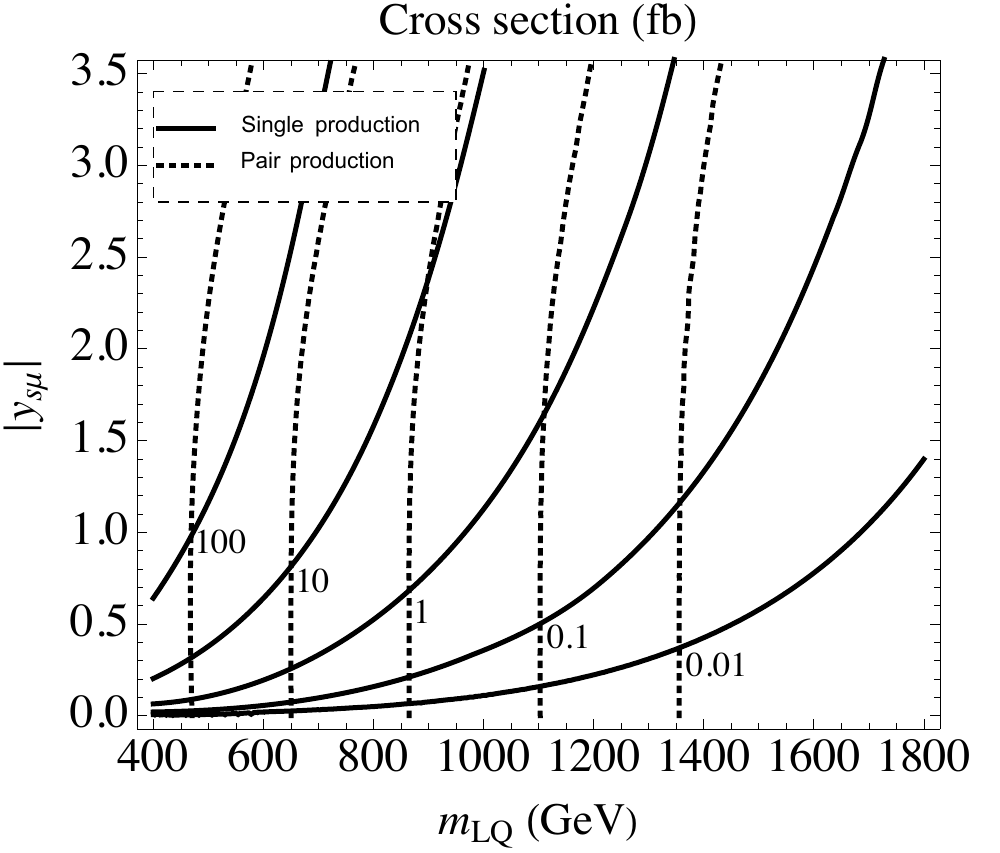}
\caption{\label{fig:-plot-1} Contours of constant leading-order cross sections for single leptoquark production (solid lines)
and the leptoquark pair production (dotted lines) at $\sqrt{s}=8$\,TeV center of mass
energy in proton-proton collisions in the $(m_\mathrm{LQ},|y_i|)$ planes. The results are obtained using MadGraph and 
setting factorization and renormalization scales to $\mu_F=\mu_R=m_\mathrm{LQ}/2$. 
The region shown in grey is excluded from the atomic parity violation constraints.
See the text for the details.}
\end{figure} 
The final results are shown in figure~\ref{fig:-plot-1} in terms of contours of constant cross 
section in the $(m_\mathrm{LQ},|y_i|)$ plane. We give separate predictions for $y_{ue}$, $y_{de}$ and $y_{s\mu}$ couplings. To avoid any potential confusion we explicitly write $y_1=y_{ue}$ in figure~\ref{fig:-plot-1} when we plot the impact of Yukawa coupling of $R_2^{5/3}$ with an electron and an up-quark on total cross section. In case of $\tilde{R}_2^{2/3}$ we separately consider contributions from $y_1=y_{de}$ and $y_2=y_{s \mu}$. The contours of constant cross 
section in figure~\ref{fig:-plot-1} for the single leptoquark production are shown in solid lines, while the
contours for the leptoquark pair production are shown in dotted lines.

The leptoquark pair production is clearly fixed by the QCD 
pair production for small Yukawa couplings. This behavior corresponds to a region where dotted lines run vertically in figure~\ref{fig:-plot-1}. However, in the large coupling regime, the total cross section can be significantly enhanced by the $t$-channel contribution. (Note the negative interference between the $s$- and the $t$-channels in a transition region.) The leptoquark single production, on the other hand, vanishes completely in the zero coupling limit. However, due to the final state phase space, it drops less rapidly with larger LQ masses compared to the LQ pair production. In other words, the contributions from this production mechanism
become increasingly important at larger LQ masses. Note that the relative strengths of two cross sections depend on the parton distribution functions of the initial
state partons. The largest (smallest) effect for the fixed value of appropriate Yukawa coupling is seen for the $y_{u e}$ ($y_{s \mu}$) case. 

To sum up, the contributions from the additional production mechanism seem to 
have dramatic impact on the leptoquark phenomenology at the LHC. Although the
conclusions made here are based solely on the cross section calculations we show that this also holds at the analysis level in section~\ref{sec:recast}. Interestingly enough, the recast of the existing CMS search for the second generation leptoquarks yields an improved constraint on the LQ parameter space. Such analysis would be even more important for the first generation leptoquarks in view of preceding discussion. 

Before we present the recast, it remains to be seen whether large Yukawa couplings are allowed by existing flavor physics measurements. We discuss this question in the next section. 

\section{Flavor constraints}
\label{sec:flavor}

We pursue a simplified scenario where only one diagonal element in $y$ for either $\tilde{R}_2^{2/3}$ or $R_2^{5/3}$ dominates. We now show that this approach is consistent with constraints from flavor physics. We furthermore demonstrate that this is the only viable scenario whenever one investigates regime of large Yukawa couplings for $\tilde{R}_2^{2/3}$ leptoquark.

The leptoquark production mechanism effects we discuss in section~\ref{sec:production} do not involve $y_3$ coupling at all due to the particularities of the proton composition. We thus neglect it in what follows and only investigate constraints on $|y_1|$, $|y_2|$ and a $|y_1 y^*_2|$ product for $\tilde{R}_2^{2/3}$ and $R_2^{5/3}$ as a function of the LQ mass. We refer to $y_1$ and $y_2$ of $\tilde{R}_2^{2/3}$ ($R_2^{5/3}$) as $y_{de}$ and $y_{s \mu}$ ($y_{ue}$ and $y_{c \mu}$), respectively. 

There exists a meaningful upper bound on $|y_1|/m_\mathrm{LQ}$ from atomic parity violation (APV) experiments, whereas a tight constraint on $|y_{de} y^*_{s\mu}|/m_\mathrm{LQ}^2$ arises from $K_L \rightarrow \mu^- e^+$ processes. A constraint on $|y_{ue} y^*_{c\mu}|/m_\mathrm{LQ}^2$ originates from $D^0 \rightarrow \mu^- e^+$ and is rather weak. We also find $|y_{s \mu}|$ not to be constrained by any experimental data from flavor physics including $g-2$ of muon. Coupling $|y_{c \mu}|$ of $R_2$, on the other hand, is slightly constrained by data on $g-2$ of muon but it can still be of the order of unity for experimentally viable leptoquark masses. We discuss these constraints in what follows in more detail.

\subsection{Atomic parity violation (APV)}

The effective Lagrangian leading to APV, below electroweak scale,  can be written as in ref.~\cite{Gresham:2012wc}: 
 \begin{equation}
{\cal L}_\mathrm{PV} = \frac{G_F}{\sqrt 2} \sum_{q=u,d} ( C_{1q} \bar e \gamma^\mu \gamma_5 e \bar q \gamma_\mu q + C_{2q} \bar e \gamma^\mu e \bar q \gamma_\mu  \gamma_5 q )\, .
\label{e1PV}
\end{equation}
Within the SM the $Z^0$ boson exchange leads to the following coefficients: $C^\mathrm{SM}_{1u} = -1/2 +4/3\, \sin^2 \theta_W$ and $C^\mathrm{SM}_{1d} = 1/2 -2/3\, \sin^2 \theta_W$. The higher-order corrections within the SM are determined in refs.~\cite{Marciano:1982mm,RamseyMusolf:1999qk} enabling one to generate very precise constraints on the potential contributions from new physics. APV is dependent on the nuclear weak charge defined as $Q_W( Z,N)= -2 [(2 Z+N) C_{1u} +(2N +Z) C_{1d}]$~\cite{Gresham:2012wc}, where $C_{1u}=C^\mathrm{SM}_{1u}+\delta C_{1u}$ and $C_{1d}=C^\mathrm{SM}_{1d}+\delta C_{1d}$. $\delta C_{1u}$ ($\delta C_{1u}$) is the new physics contribution generated by the presence of $R_2^{5/3}$ ($\tilde{R}_2^{2/3}$ and/or $R_2^{2/3}$). Here, $Z$ represents atomic number and $N$ stands for neutron number. 

The experimentally extracted value $Q_W (\mathrm{Cs})= -73.20(35)$ for the  cesium atom $( ^{133}\mathrm{Cs})$ \cite{Wood:1997zq,Guena:2004sq} is in very good agreement with the SM result $Q_W (\mathrm{Cs})= -73.15(35)$~\cite{Porsev:2009pr}. This gives a tight constraint on the effective coefficients $\delta C_{1u}$ and $\delta C_{1d}$ that, for the LQ contribution, reads
 \begin{equation}
 \delta C_{1 u (d)}= \frac{\sqrt 2}{G_F} \frac{|y_{u (d)e}|^2}{8 m_\mathrm{LQ}^2}.
 \label{e2PV}
 \end{equation}
This translates into the following limits on $|y_{de} |$ and $|y_{ue} |$ if one requires a $2\,\sigma$ agreement with the experimental measurement of $Q_W (\mathrm{Cs})$:
 \begin{equation}
 |y_{de} | \leq 0.34 \left(\frac{m_\mathrm{LQ}}{1\,\mathrm{TeV}}\right)\,, \qquad |y_{ue} | \leq 0.36 \left(\frac{m_\mathrm{LQ}}{1\,\mathrm{TeV}}\right).
 \label{e3PV}
\end{equation}
The bounds presented in eq.~\eqref{e3PV} are extracted under the assumption that only one of the two contributions is present at a given moment. This assumption cannot be realised if one considers $R_2$ leptoquark and takes $z_{11} \neq 0$. (See eq.~\eqref{eq:L23} for details.) Note, however, that it is not possible to cancel $\delta C_{1u}$ against $\delta C_{1d}$ or vice versa. This means that the upper bounds presented in eq.~\eqref{e3PV} are applicable in the most general case. We accordingly use these bounds to exclude shaded regions in two upper panels in figure~\ref{fig:-plot-1}.
  
\subsection{\texorpdfstring{$K_L \rightarrow \mu^- e^+$}{}}

The diagonal couplings of $\tilde{R}_2^{2/3}$ enter at the tree level into the lepton flavor violating $K_L \to \mu^{-} e^{+}$ decay amplitude. Following ref.~\cite{Dorsner:2011ai} one can write down the decay width:
\begin{equation}
\Gamma_{K_L \to \mu^{-} e^{+}} = \frac{ |y_{s\mu} y^*_{d e}|^2}{512 \pi} \frac{m_K^3 f_K^2}{m_\mathrm{LQ}^4} \left(\frac{m_\mu}{m_K}\right)^2 \left[1-  \left(\frac{m_\mu}{m_K}\right)^2 \right]^2\,.
\label{e1KLFV}
\end{equation}
Using lattice QCD result $f_K =156.1 (0.8)$\,MeV~\cite{Aoki:2013ldr} and $BR(K_L\to \mu^\pm e^\mp) < 4.7 \times 10^{-12}$~\cite{Beringer:1900zz}, we derive the following bound 
\begin{equation}
|y_{s \mu} y^*_{d e}|< 2.1  \times 10^{-5} \left(\frac{m_\mathrm{LQ}}{1 {\rm TeV}}\right)^2.
\label{e2KLFV}
\end{equation}

\subsection{\texorpdfstring{$D^0 \rightarrow \mu^- e^+$}{}}

The diagonal couplings of $R_2^{5/3}$ to the SM fermions enter at the tree level into the lepton flavor violating $D^0 \to \mu^{-} e^{+}$ decay amplitude. The decay width reads
\begin{equation}
\Gamma_{D^0 \to \mu^{+} e^{-}} = \frac{ |y_{c\mu} y^*_{u e}|^2}{256 \pi} \frac{m_D^3 f_D^2}{m_\mathrm{LQ}^4} \left(\frac{m_\mu}{m_D}\right)^2 \left[1-  \left(\frac{m_\mu}{m_D}\right)^2 \right]^2\,.
\label{e1DLFV}
\end{equation}
Using lattice QCD result $f_D =209.2$\,MeV~\cite{Na:2012iu} and taking $BR(D^0\to \mu^\pm e^\mp) < 2.6 \times 10^{-7}$~\cite{Beringer:1900zz}, we find the following bound 
\begin{equation}
|y_{c \mu} y^*_{u e}|< 0.6 \left(\frac{m_\mathrm{LQ}}{1 {\rm TeV}}\right)^2.
\label{e2DLFV}
\end{equation}

\subsection{\texorpdfstring{$g-2$}{g-2} of muon}

The $\tilde{R}_2^{2/3}$ coupling $y_{s\mu}$ can, in principle, contribute to the muon $g-2$ with a leptoquark and a strange quark within the loop. However, following refs.~\cite{Cheung:2001ip,Dorsner:2013tla,Dorsner:2011ai,Queiroz:2014zfa}, it is easy to show that the muon $g-2$ anomaly does not constrain $y_{s\mu}$ at all. This is due to a smallness of the strange quark mass and a substantial cancellations of the two relevant contributions that enter into a shift of the muon anomalous moment with respect to the SM value. The electric charge of $R_2^{5/3}$, on the other hand, does not allow for the aforementioned cancellation. The consequence of that is a mild constraint on $|y_{c\mu}|/m_\mathrm{LQ}$ that reads~\cite{Dorsner:2013tla}
\begin{equation}
 |y_{c\mu} | \leq 1.0 \left(\frac{m_\mathrm{LQ}}{1\,\mathrm{TeV}}\right).
 \label{e1g-2}
\end{equation}

The preceding discussion basically demonstrates that if one takes $y_{s \mu}$ to be large, i.e., an order one quantity, then $y_{d e}$ has to be very small in order to satisfy eq.~\eqref{e2KLFV}. (The situation with the $R_2^{5/3}$ couplings is somewhat more involved. The flavor physics constraints allow for $y_{u e}$ and $y_{c \mu}$ to simultaneously be of relatively large value.) This simply means that we can consistently set to zero $y_{d e}$ as we concentrate on the study of the effects of large $y_{s \mu}$. Our recast of the second generation leptoquark search by the CMS collaboration is thus self consistent and well justified in that regime. We turn to it in the next section.
 
\section{Recasting the CMS search for the second generation leptoquarks}
\label{sec:recast}

The CMS collaboration has recently reported a search for the second generation 
scalar leptoquarks based on $19.6$\,fb$^{-1}$ of data at $\sqrt{s}=8$\,TeV proton-proton center of mass energy~\cite{CMS:zva}. 
The underling assumption is that the corresponding $\mu$--$s$--$LQ$ coupling, i.e., $y_{s \mu}$, is small. The LQ pair production is thus completely fixed by QCD as discussed before. Recalling 
the results of section~\ref{sec:production}, we relax this assumption and study the impact 
of large $y_{s \mu}$ coupling on the existing experimental search. In particular, large coupling
leads to substantial signal yield from $t$-channel leptoquark pair production as well 
as single leptoquark production. This, then, provides more restrictive constraint on 
the LQ parameter space. In the following, we recast the CMS search reported in~\cite{CMS:zva}
in order to set an improved limit on the second generation scalar leptoquark parameter space.

Here we outline the details of our analysis. We use FeynRules
(v1.6.16)~\cite{Christensen:2008py} to implement the model containing $(\mathbf{3},\mathbf{2},1/6)$ scalar representation and the interactions defined by the 
Lagrangian in eq.~\eqref{eq:main_1}. We use MadGraph~5 (v1.5.11)~\cite{Alwall:2011uj} 
to generate $p\,p \to \tilde{R}_{2}^{2/3}\,\tilde{R}_{2}^{2/3\,*}$, 
$p\,p \to \tilde{R}_{2}^{2/3\,*}\,\mu^+$ and $p\,p \to \tilde{R}_{2}^{2/3}\,\mu^-$ processes, followed 
by the leptoquark decays to a muon and a strange quark. The decay branching ratio 
is taken to be $\beta_2=1$, which is in line with our assumption of a single large coupling and is also consistent with the flavor physics constraints we present in section~\ref{sec:flavor}. 
The next-to-leading order QCD corrections to the LQ pair production are shown to substantially enhance 
the tree level cross section~\cite{Kramer:2004df}. However, the analysis conducted here is based 
on LO calculations. This makes exclusion limits we present conservative. In order to 
partially account for large corrections, we fix the factorization and 
renormalization scales to $\mu_F=\mu_R=m_\mathrm{LQ}/2$. 
We simulate showering and hadronization 
effects using Pythia (v6.426)~\cite{Sjostrand:2006za}. As a detector
 simulator we chose the default implementation of the 
CMS detector in Delphes~(v3.0.9)~\cite{deFavereau:2013fsa}. 
In addition, we have modified the default implementation by switching to the anti-$k_T$ jet algorithm with distance 
parameter $R=0.5$, and by changing the muon isolation criteria in accordance with ref.~\cite{CMS:zva}.

We adopt the following preselection cuts~\cite{CMS:zva}:
\begin{itemize}
\item We require at least two muons with $p_T > 45$\,GeV and $|\eta|<2.1$. Two muons with the highest 
	$p_T$ are required to have 
	spatial separation $\Delta R > 0.3$ and invariant mass $>50$\,GeV;
\item We require at least two jets with $p_T > 45$\,GeV and $|\eta|<2.4$ and $\Delta R >0.3$ spatial separation
	from muon candidates. The leading jet $p_T$ is required to be $>125$\,GeV;
\item The scalar sum of the transverse momenta ($S_T$) of two leading $p_T$ muons and two leading $p_T$ jets 
	is required to be $S_T>300$\,GeV.
\end{itemize}

\begin{table}[tbp]
\centering
\begin{tabular}{|c|ccccc|}
\hline
$m_\mathrm{LQ}\,$(GeV) & 500  & 700 & 900 & 950 & $\geq1000$\\
\hline 
$S_{T}\,>$(GeV) & $685$   & $935$  & $1135$ & $1175$ & $1210$\\
$M_{\mu\mu}\,> $(GeV) & $150$  & $195$ & $230$ & $235$ & $245$\\
$M_\mathrm{min}(\mu,j)\,> $(GeV) & $155$   & $295$ & $535$ & $610$ & $690$\\
\hline 
\hline 
Signal yield $<$ at $95\%\,$CL & $34$  & $9.8$  & $5.6$ & $3.5$ & $1.8$\\
\hline
\end{tabular}
\caption{\label{tab:cuts} Final selection cuts as used by the CMS collaboration. The cuts are optimized 
		for the QCD pair production and depend on the leptoquark mass
		hypothesis. The last raw shows the observed $95\%\,$CL upper limit on 
		the allowed signal yield after final selection cuts are applied~\cite{CMS:zva}.}
\end{table}

\begin{figure}[tbp]
\centering
\includegraphics[scale=0.80]{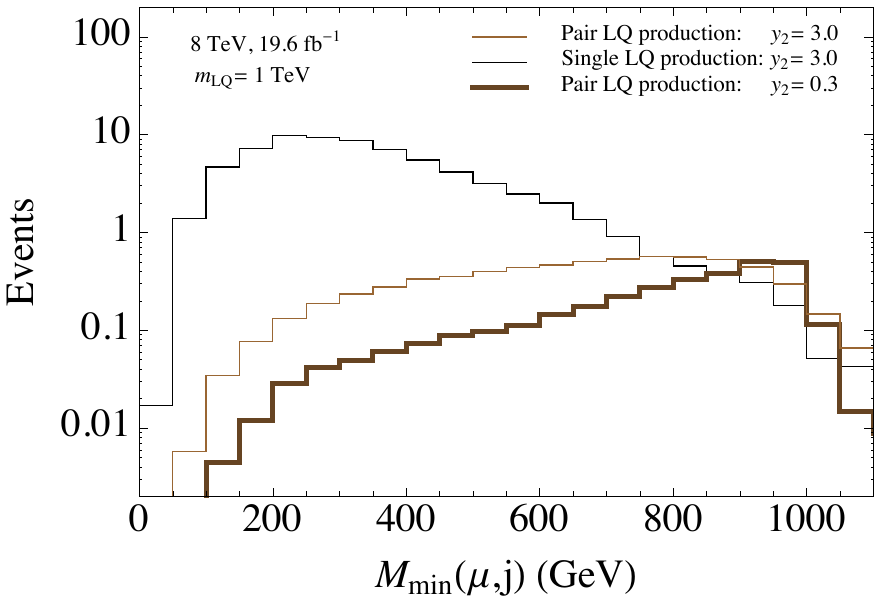} \hfill
\includegraphics[scale=0.80]{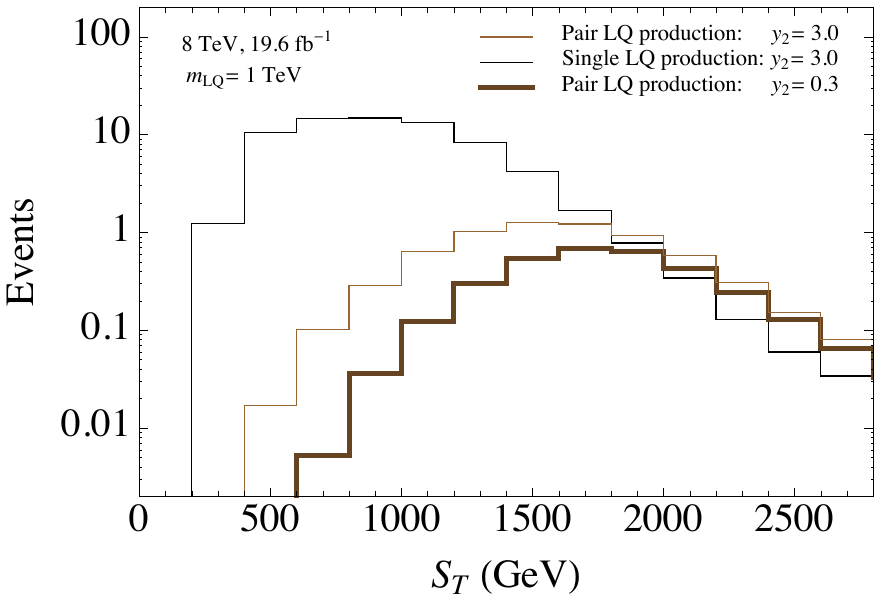} 
\caption{\label{fig:-plot-2} The signal events distributions in $M_\mathrm{min}(\mu,j)$ and $S_T$ variables 
		after preselection
		cuts applied only. The predictions for single LQ production for $y_2=3$ are
		shown in black thin, while the predictions for pair production with the 
		same value of the coupling
		are shown in light brown thin. The predictions due to the QCD pair 
		production are shown in dark brown thick. The leptoquark mass is taken to be
		$m_\mathrm{LQ}=1$\,TeV.}
\end{figure} 

The final selection cuts are applied on the following three variables: (i) the invariant mass 
of the dimuon pair ($M_{\mu\mu}$), (ii) the scalar sum of the transverse momenta of 
the two leading $p_T$ muons and the two leading $p_T$ jets ($S_T$) and (iii) the 
smallest of the two muon-jet invariant masses that minimizes the leptoquark mass difference
($M_\mathrm{min}(\mu,j)$). The final cuts used by the CMS collaboration are reported in table~\ref{tab:cuts}. These are optimized for the QCD pair production and depend 
on the leptoquark mass hypothesis. In order to illustrate the impact of the final
selection cuts in the large coupling regime we plot in figure~\ref{fig:-plot-2} the 
signal events distributions in $S_T$ and $M_\mathrm{min}(\mu,j)$ variables after the application of preselection cuts only. Here, we choose leptoquark mass to be $m_\mathrm{LQ}=1$\,TeV which is close to a present exclusion limit.
The signal yield from the QCD pair production is show in dark-brown thick line. As expected, the distributions in $S_T$ variable tend to peak for $S_T \sim 2 m_\mathrm{LQ}$, while the
distributions in $M_\mathrm{min}(\mu,j)$ variable peak at $M_\mathrm{min}(\mu,j)\sim m_\mathrm{LQ}$.
In light-brow thin lines, we show the contribution due to leptoquark pair 
production for the value of the coupling set to $y_2=3.0$. The integrated signal
yield is larger compared to the previous case due to additional contributions from the 
$t$-channel diagrams. Unfortunately, the events tend to populate slightly lower 
$S_T$ and $M_\mathrm{min}(\mu,j)$ parameter regions. Finally, we show the contributions
from a single leptoquark production for the coupling $y_2=3.0$ in black thin lines.
While the integrated signal yield is significantly larger with respect to previous 
cases, the events tend to have considerable smaller $S_T$ and $M_\mathrm{min}(\mu,j)$
values. Obviously, the variable $M_\mathrm{min}(\mu, j)$ is efficient only in the presence 
of leptoquark pair decaying to a final state particles. Furthermore, we have checked 
that the muon coming from the production tends to have considerable smaller 
$p_T$ with respect to the muon coming from the leptoquark decay. The same holds 
for the second leading $p_T$ jet, likely originating from the real QCD radiation, 
compared to the leading $p_T$ jet, likely coming from the decay. 
This explains the softer distributions in the $S_T$ variable. 
Conclusively, applying the appropriate final selection cuts from table~\ref{tab:cuts},
the majority of signal is lost. We do not aim here to optimise the search 
for the single LQ production, but rather to illustrate the importance of its inclusion.  
We thus keep the cuts as used by the CMS collaboration in order to rely on the official background predictions.

\begin{table}[tbp]
\centering
\begin{tabular}{|c|cccc|}
\hline
\multirow{2}{*}{$m_\mathrm{LQ}(\textrm{GeV})$} & \multicolumn{4}{c|}{$N_\mathrm{evs}(\textrm{Pair production})+N_\mathrm{evs}(\textrm{Single production})$ }\\
 & $y_{2}=0.3$ & $y_{2}=1.0$ & $y_{2}=2.0$ & $y_{2}=3.0$\\
 \hline 
500 & $\bf 600+8.2$ & $\bf 600+89$ & $\bf 720+330$ & $\bf 1300+700$\\
700 & $\bf 55+0.98$ & $\bf 56+11$ & $\bf 64+41$ & $\bf 110+81$\\
900 & $\bf 6.5+0.10$ & $\bf 6.5+1.2$ & $\bf 7.0+4.5$ & $\bf 11+8.4$\\
1000 & $\bf 2.2+0.03$ & $\bf 2.2+0.33$ & $\bf 2.3+1.1$ & $\bf 3.1+2.3$\\
1050 & $1.5+0.02$ & $\bf 1.5+0.27$ & $\bf 1.5+1.0$ & $\bf 2.1+2.1$\\
1100  & $0.96+0.02$ & $0.96+0.21$ & $\bf 1.0+0.82$ & $\bf 1.4+1.6$\\
1150  & $0.62+0.02$& $0.62+0.17$ & $0.66+0.75$ & $\bf 0.92+1.4$\\
1200 & $0.41+0.01$ & $0.41+0.14$ & $0.44+0.55$ & $\bf 0.60+1.3$\\
1300 & $0.17+0.01$ & $0.17+0.09$ & $0.19+0.37$ & $0.26+0.74$  \\
1400 & $0.07+0.00$ & $0.07+0.06$ & $0.08+0.24$ & $0.12+0.52$  \\
\hline
\end{tabular}
\caption{\label{tab:cms} The signal yields obtained from the simulation after final selection 
		cuts are applied. The predictions are shown as the sum of the pair production
		and the single production contributions, respectively. The points in parameter space spanned by $y_2\equiv y_{s \mu}$ and $m_\mathrm{LQ}$ shown in bold
		are excluded by the existing data. See the text for the details.}
\end{table}

We present the results of our simulation in table~\ref{tab:cms}. In particular,
we show the signal event yields for certain choices of coupling $y_2 \equiv y_{s \mu}$ and leptoquark
mass $m_\mathrm{LQ}$. The predictions are given as the sum of two numbers 
representing the individual contributions from the leptoquark pair and single 
productions, respectively.
We have checked that our predictions for the LQ pair production for small couplings 
agree well with the results reported in table 4 of~\cite{CMS:zva} after using the next-to-leading order QCD corrected cross sections. This serves as an important cross-check of our 
simulation procedure.
The main observation at this point is that, for large values of the coupling $y_2$, 
the contributions from the $t$-channel pair production and, particularly, single LQ
production become important. Moreover, the latter process is especially relevant
for larger LQ masses due to the phase space suppression in pair production as 
discussed before. The point made here gains on importance as our recast sets stronger exclusion limits on LQ parameter space.
 
We finally translate the predictions for signal yields from table~\ref{tab:cms}  
into exclusion regions in $(m_\mathrm{LQ}, y_{s \mu})$ plane.  Here we rely on the official statistical
analysis performed by the CMS collaboration. 
In particular, the observed $95\%$ CL upper limits on the allowed signal yields, after
 final selection cuts are applied, are shown in the last raw of table~\ref{tab:cuts}. 
These are obtained by rescaling the observed $95\%$ CL upper limits on the 
production cross sections as reported in figure~8 of~\cite{CMS:zva}. 
The rescaling factors are the signal event yields reported in table 4 
of~\cite{CMS:zva} divided by the SM cross  section from table 1 of~\cite{CMS:zva}.

\begin{figure}[tbp]
\centering
\includegraphics[scale=1.0]{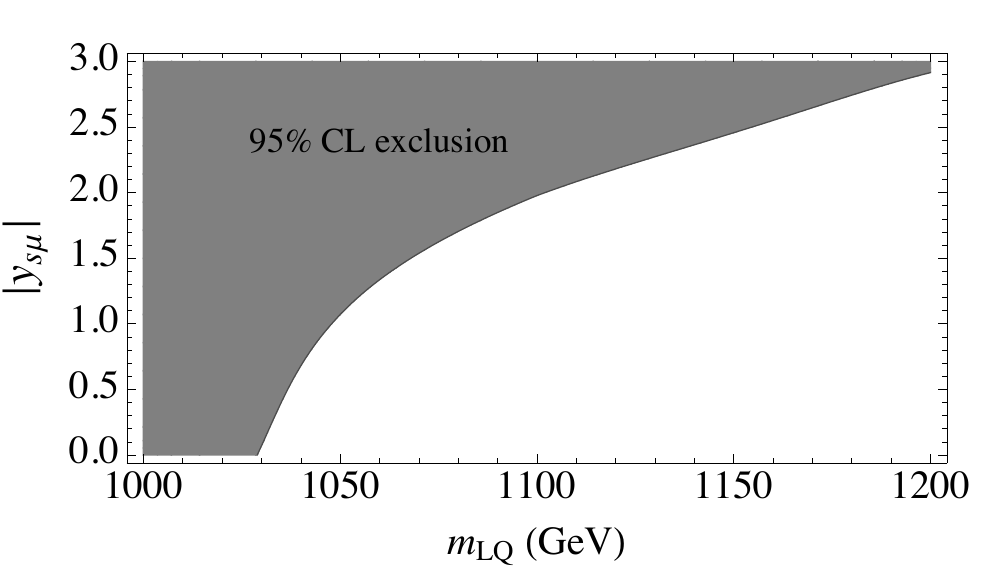}
\caption{\label{fig:-plot-3} The improved direct contraints on the second generation scalar leptoquarks
		 from the existing LHC data. The region shown in grey is excluded at $95\%$ 				CL.}
\end{figure}

The improved constraint on the second generation leptoquark parameter space is shown
in figure~\ref{fig:-plot-3}. The parameter region in $(m_\mathrm{LQ}, y_{s \mu})$ plane
excluded at $95\%$ CL by the existing LHC data is shown in grey. The signal
yield as a function of the coupling and mass, i.e., $N_\mathrm{evs}(m_\mathrm{LQ},y_{s \mu})$, is obtained
after interpolating over the points shown in table~\ref{tab:cms}. The excluded
region corresponds to the points for which $N_\mathrm{evs}(m_\mathrm{LQ},y_{s \mu})$ is greater than 
the appropriate value reported in table~\ref{tab:cuts}. 
As advocated before, the limits on the LQ masses are more stringent for larger values of $\mu$--$s$--$LQ$ coupling.

\section{Conclusions}
\label{sec:conclusions}

We study a pair production and a single production of leptoquarks in the regime when leptoquarks couple strongly to a charged lepton and a quark of the same generation. This we do for two particular leptoquarks that do not cause proton decay at tree level. We also discuss existing flavor constraints on the strength of the relevant Yukawa couplings and show that the regime we are interested in is viable with respect to experimental measurements. 

We demonstrate the importance of inclusion of the single leptoquark production and the $t$-channel pair production through a recast of an existing CMS search at LHC for the second generation leptoquark that couples to a muon and a strange quark. The recast yields the best limit on the Yukawa coupling and mass of the second generation leptoquark to date.

As the exclusion limits on the leptoquark masses approach TeV scale, the relative importance of the single leptoquark production over the pair production significantly increases. We therefore strongly suggest future experimental searches along these lines.

\acknowledgments
This work has been supported in part by Croatian Science Foundation under the project 7118. I.D.\ acknowledges the SNSF support through the SCOPES project No.\ IZ74Z0\_137346. 
S.F.\ and A.G.\ acknowledge the support by the Slovenian Research Agency (ARRS). We thank J.F.\ Kamenik, V.\ Brigljevi\' c and U.\ Langenegger for insightful discussions.


\begin{thebibliography}{99}  

\bibitem{Pati:1974yy} 
  J.~C.~Pati and A.~Salam,
  Phys.\ Rev.\ D {\bf 10}, 275 (1974)
  [Erratum-ibid.\ D {\bf 11}, 703 (1975)].

\bibitem{Georgi:1974sy} 
  H.~Georgi and S.~L.~Glashow,
  Phys.\ Rev.\ Lett.\  {\bf 32}, 438 (1974).
  
\bibitem{Buchmuller:1986zs} 
  W.~Buchmuller, R.~Ruckl and D.~Wyler,
  Phys.\ Lett.\ B {\bf 191}, 442 (1987)
  [Erratum-ibid.\ B {\bf 448}, 320 (1999)].

\bibitem{Hewett:1987yg} 
 J.~L.~Hewett and S.~Pakvasa,
 Phys.\ Rev.\ D {\bf 37}, 3165 (1988).

\bibitem{Eboli:1987vb} 
  O.~J.~P.~Eboli and A.~V.~Olinto,
  Phys.\ Rev.\ D {\bf 38}, 3461 (1988).
  
\bibitem{DeMontigny:1989yd} 
 M.~De Montigny and L.~Marleau,
 Phys.\ Rev.\ D {\bf 40}, 2869 (1989)
 [Erratum-ibid.\ D {\bf 56}, 3156 (1997)].

\bibitem{Ohnemus:1994xf} 
  J.~Ohnemus, S.~Rudaz, T.~F.~Walsh and P.~M.~Zerwas,
  Phys.\ Lett.\ B {\bf 334}, 203 (1994)
  [hep-ph/9406235].
  
\bibitem{Eboli:1999ye} 
  O.~J.~P.~Eboli and T.~L.~Lungov,
  Phys.\ Rev.\ D {\bf 61}, 075015 (2000)
  [hep-ph/9911292].
    
\bibitem{Belyaev:2005ew} 
  A.~Belyaev, C.~Leroy, R.~Mehdiyev and A.~Pukhov,
  JHEP {\bf 0509}, 005 (2005)
  [hep-ph/0502067].
  
\bibitem{Dorsner:2012nq} 
  I.~Dorsner, S.~Fajfer and N.~Kosnik,
  Phys.\ Rev.\ D {\bf 86}, 015013 (2012)
  [arXiv:1204.0674 [hep-ph]].

\bibitem{Davidson:2010uu} 
  S.~Davidson and S.~Descotes-Genon,
  JHEP {\bf 1011}, 073 (2010)
  [arXiv:1009.1998 [hep-ph]].

\bibitem{Aaron:2011zz} 
  F.~D.~Aaron {\it et al.}  [H1 Collaboration],
  Phys.\ Lett.\ B {\bf 701}, 20 (2011)
  [arXiv:1103.4938 [hep-ex]].

\bibitem{Alwall:2011uj} 
  J.~Alwall, M.~Herquet, F.~Maltoni, O.~Mattelaer and T.~Stelzer,
  JHEP {\bf 1106}, 128 (2011)
  [arXiv:1106.0522 [hep-ph]].
        
\bibitem{Christensen:2008py} 
  N.~D.~Christensen and C.~Duhr,
  Comput.\ Phys.\ Commun.\  {\bf 180}, 1614 (2009)
  [arXiv:0806.4194 [hep-ph]].

\bibitem{Gresham:2012wc}
  M.~I.~Gresham, I.~-W.~Kim, S.~Tulin and K.~M.~Zurek,
  Phys.\ Rev.\ D {\bf 86} (2012) 034029
  [arXiv:1203.1320 [hep-ph]].

\bibitem{RamseyMusolf:1999qk}
  M.~J.~Ramsey-Musolf,
  Phys.\ Rev.\ C {\bf 60} (1999) 015501
  [hep-ph/9903264].

\bibitem{Marciano:1982mm}
  W.~J.~Marciano and A.~Sirlin,
  Phys.\ Rev.\ D {\bf 27} (1983) 552.

\bibitem{Wood:1997zq}
  C.~S.~Wood, S.~C.~Bennett, D.~Cho, B.~P.~Masterson, J.~L.~Roberts, C.~E.~Tanner and C.~E.~Wieman,
  Science {\bf 275} (1997) 1759.
 
\bibitem{Guena:2004sq}
  J.~Guena, M.~Lintz and M.~A.~Bouchiat,
  Phys.\ Rev.\ A {\bf 71} (2005) 042108
  [physics/0412017 [physics.atom-ph]].

\bibitem{Porsev:2009pr}
  S.~G.~Porsev, K.~Beloy and A.~Derevianko,
  Phys.\ Rev.\ Lett.\  {\bf 102} (2009) 181601
  [arXiv:0902.0335 [hep-ph]].

\bibitem{Dorsner:2011ai}
  I.~Dorsner, J.~Drobnak, S.~Fajfer, J.~F.~Kamenik and N.~Kosnik,
  JHEP {\bf 1111} (2011) 002
  [arXiv:1107.5393 [hep-ph]].

\bibitem{Aoki:2013ldr}
  S.~Aoki, Y.~Aoki, C.~Bernard, T.~Blum, G.~Colangelo, M.~Della Morte, S.~D\"{u}rr and A.~X.~El Khadra {\it et al.},
  arXiv:1310.8555 [hep-lat].

\bibitem{Beringer:1900zz}
  J.~Beringer {\it et al.}  [Particle Data Group Collaboration],
  Phys.\ Rev.\ D {\bf 86} (2012) 010001.

\bibitem{Na:2012iu} 
  H.~Na, C.~T.~H.~Davies, E.~Follana, G.~P.~Lepage and J.~Shigemitsu,
  Phys.\ Rev.\ D {\bf 86}, 054510 (2012)
  [arXiv:1206.4936 [hep-lat]].

\bibitem{Cheung:2001ip} 
  K.~-m.~Cheung,
  Phys.\ Rev.\ D {\bf 64}, 033001 (2001)
  [hep-ph/0102238].

\bibitem{Dorsner:2013tla} 
  I.~Dor\v sner, S.~Fajfer, N.~Ko\v snik and I.~Ni\v sand\v zi\' c,
  JHEP {\bf 1311}, 084 (2013)
  [arXiv:1306.6493 [hep-ph]].

\bibitem{Queiroz:2014zfa} 
  F.~S.~Queiroz and W.~Shepherd,
  Phys.\ Rev.\ D {\bf 89}, 095024 (2014)
  [arXiv:1403.2309 [hep-ph]].
  
\bibitem{CMS:zva} 
  [CMS Collaboration],
  CMS-PAS-EXO-12-042.

\bibitem{Kramer:2004df} 
  M.~Kramer, T.~Plehn, M.~Spira and P.~M.~Zerwas,
  Phys.\ Rev.\ D {\bf 71}, 057503 (2005)
  [hep-ph/0411038].
  
\bibitem{Sjostrand:2006za} 
  T.~Sjostrand, S.~Mrenna and P.~Z.~Skands,
  JHEP {\bf 0605}, 026 (2006)
  [hep-ph/0603175].

\bibitem{deFavereau:2013fsa} 
  J.~de Favereau {\it et al.}  [DELPHES 3 Collaboration],
  JHEP {\bf 1402}, 057 (2014)
  [arXiv:1307.6346 [hep-ex]].
   
\end{thebibliography}
\end{document}